\documentclass[12pt,epsfig,aps]{article}
\usepackage{graphicx}
\usepackage{epsfig}
\usepackage{amssymb,latexsym,amsmath}
\usepackage{color}
\newcommand{\bs}{\mathbf}

\begin{document}
\title{ \Large{ \bf Pionic Freeze-out Hypersurfaces in Relativistic
                    Nucleus-Nucleus Collisions} }
\author{
D. Anchishkin$^1$, V. Vovchenko$^2$, L.P. Csernai$^3$
\\ $^1${\small \it Bogolyubov Institute for Theoretical Physics,
             03680 Kiev, Ukraine}
\\ $^2${\small \it Taras Shevchenko Kiev National University,
03022 Kiev, Ukraine}
\\ $^3${\small \it Institute for Physics and Technology, University of Bergen,
           5007 Bergen, Norway}  }


\maketitle

\begin{abstract}
The space-time structure of the multipion system created in central
relativistic heavy-ion collisions is investigated. Using the microscopic transport model
UrQMD we determine the freeze-out hypersurface from
equation on pion density $n(t,\, \bs r)=n_c$.
It turns out that for
proper value of the critical energy density $\epsilon_c$ equation
$\epsilon(t,\, \bs r)=\epsilon_c$ gives the same freeze-out
hypersurface.
It is shown that for big enough collision energies
$E_{\rm kin} \ge 40$A~GeV/$c$ ($\sqrt{s} \ge 8$A~GeV/$c$)  the
multipion system at a time moment $\tau$ ceases to be one connected
unit but splits up into two separate spatial parts (drops), which
move in opposite directions from one another
with velocities which approach the speed of light with increase of
collision energy.
This
time $\tau$ is approximately invariant of the collision energy,
and the corresponding $\tau$=const. hypersurface can serve as
a benchmark for the freeze-out time or the transition time from
the hydrostage in hybrid models. The properties of this hypersurface
are discussed.
\end{abstract}


\maketitle

\section{Introduction}

High energy heavy ion collisions at AGS, SPS, RHIC and LHC energies are well
described by fluid dynamical models, and a substantial part of the
initial beam energy is converted into collective flow. This collective
flow is one of the most important observable phenomena arising from
the strongly interacting, high energy density quark-gluon plasma (QGP).

On the other hand the initial state, before local thermal and
chemical equilibrium is reached must be described by other means.
In the subsequent fluid dynamical (FD) stage in local equilibrium
we can take advantage of an equation of state (EoS). At the final stage
when the system becomes dilute and local thermal and
chemical equilibrium cannot be maintained we have to use another
freeze out (FO) model. Most frequently the transition from the FD
to the FO stage is assumed to happen at a 3-dimensional space-time
hypersurface where one generates the final out of equilibrium state
by using the Cooper-Frye formalism \cite{cooper-frye-PRD-v10-1974}.
When applying this formalism one should also take care for
energy, momentum and baryon charge conservation, as summarized recently
in ref. \cite{ChengEtal2010}.
This is also necessary if instead of immediately applying a
post FO distribution we supplement the FD description with
a non-equilibrium model, in a multimodule or hybrid model approach,
e.g. \cite{UrQMD1998}.
Due to the explosive dynamics of the system the FO hyper-surface has
parts both with space-like and with time-like normal vectors, which
requires a careful handling of the FO process.

In the present work we analyze the FO hyper-surface using the ultrarelativistic quantum molecular dynamics (UrQMD) model
\cite{UrQMD1998,UrQMD1999}, in the energy range of AGS,  SPS,
and RHIC for central heavy ion collisions
and the structure of these hypersurfaces under different FO conditions.
We assume in this work that the FO takes place in the hadronic phase and every
hadronic species has its own FO hypersurface.
Because pions represent the majority of the secondary particles we study first
of all the FO hypersurface for the final pion gas.

\section{Calculation algorithm}

The invariant scalar $\pi^-$ density is defined as \cite{groot}
\begin{equation}
n(x)\ =\ N^{\mu} (x) \, u_{\mu} (x) \,.
\label{eq:ninv}
\end{equation}
Here $N^{\mu}$ is the particle four-flow
\begin{equation}
N^{\mu} (x)\, =\, \int \frac{d^3 p}{p^0} \, p^{\mu} \, f(x,p)\, =\,
\left(n_{\rm lab}, n_{\rm lab}\bs v_{_E}\right) \,,
\label{eq:nmu}
\end{equation}
where $p^0=\sqrt{m_\pi^2+\bs p^2}$, $n_{\rm lab}$ is the density of negative
pions in the laboratory frame at point $x$ and $\bs v_{_E}$ is their average
three-velocity.
%
%
The quantity $u^{\mu} (x)$ is the collective four-velocity of negative pions.
It can be written as $u^{\mu} = (\gamma, \gamma \bs v)$,
where $\gamma (x) = 1/ \sqrt{1 - \bs v^2}$.
The velocity $\bs v$ is tied to Local Rest frame definition.
In Local Rest frame $u^{\mu}$ has
only temporal component: $u_{LR}^{\mu} = (1, \bs 0)$.
According to Eckart definition $u^{\mu}$ is tied to the particle (pion) flow
\begin{equation}
u^{\mu} (x)\, =\, \frac{N^{\mu}}{\left(N^{\nu} N_{\nu}\right)^{\frac{1}{2}}} \,
=\, \left(\gamma_{_E}, \gamma_{_E} \bs v_{_E}\right) \,.
\label{eq:umu}
\end{equation}
Using  eq. \eqref{eq:ninv} we can write the invariant $\pi^-$ density
according to Eckart definition as
\begin{equation}
n_{_E}(x)\ =\ \frac{n_{\rm lab}(x)}{\gamma_{_E} (x)} \,.
\end{equation}
The invariant energy density $\epsilon (x)$ is defined as
\begin{equation}
\epsilon(x)\ =\ u_{\mu} (x) \ T^{\mu \nu} (x) \ u_{\nu} (x) \,,
\label{eq:edens}
\end{equation}
where $T^{\mu \nu}$ is energy-momentum tensor
\begin{equation}
T^{\mu \nu} (x) = \int \frac{d^3 p}{p^0} \, p^{\mu} \, p^{\nu} \, f(x,p) \,.
\label{eq:Tmunu}
\end{equation}

To calculate quantities $n(x)$ and $\epsilon(x)$ we use
the transport model UrQMD v2.3 \cite{UrQMD1998,UrQMD1999}.
In order to perform calculations we take
a four-volume box with sides $L_i=20$~fm,
$i = t,x,y,z$, and
\begin{figure}[ht]
\begin{minipage}{.46\textwidth}
\centering
\includegraphics[width=\textwidth]{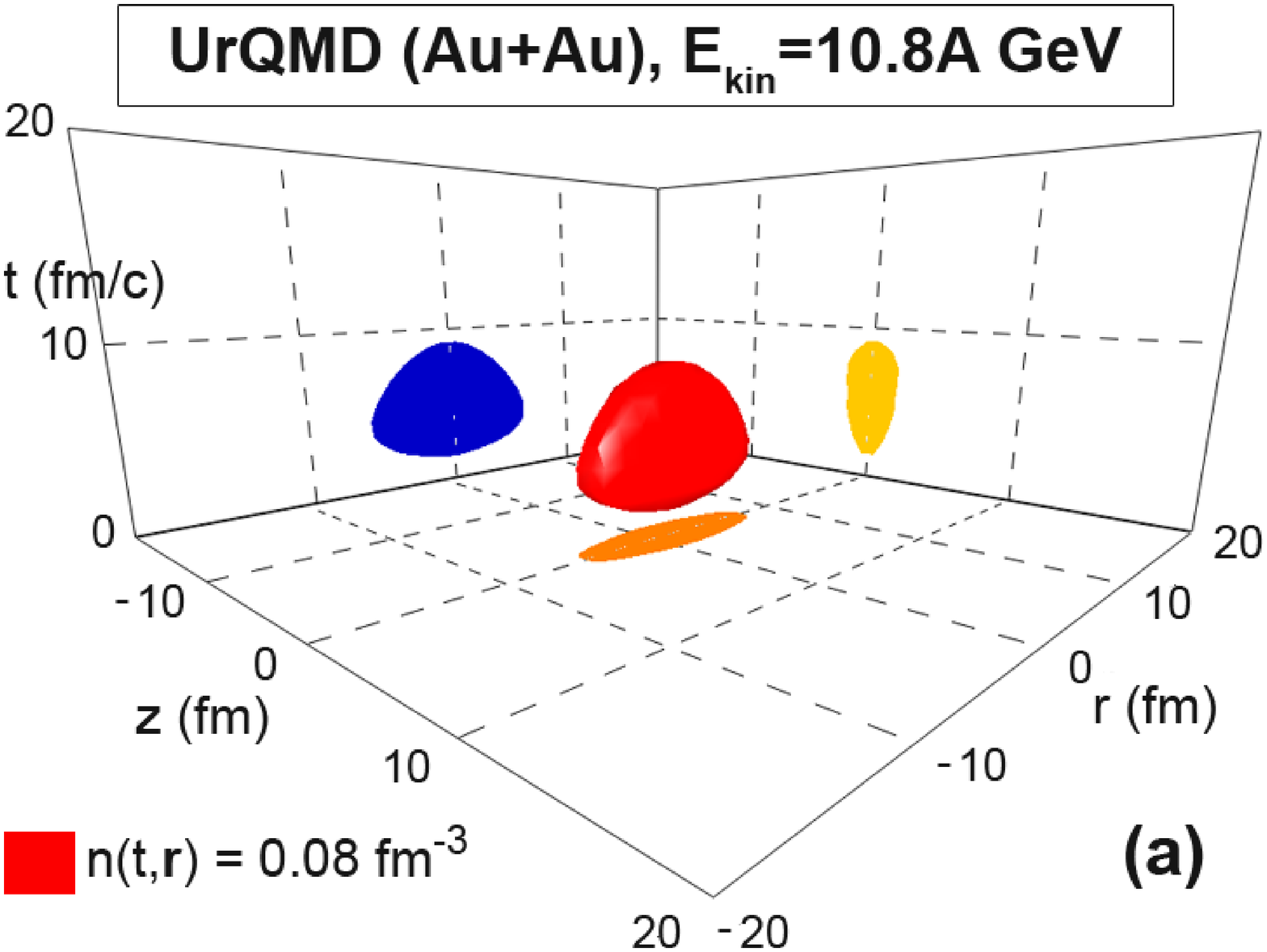}
\end{minipage}
\begin{minipage}{.46\textwidth}
\centering
\includegraphics[width=\textwidth]{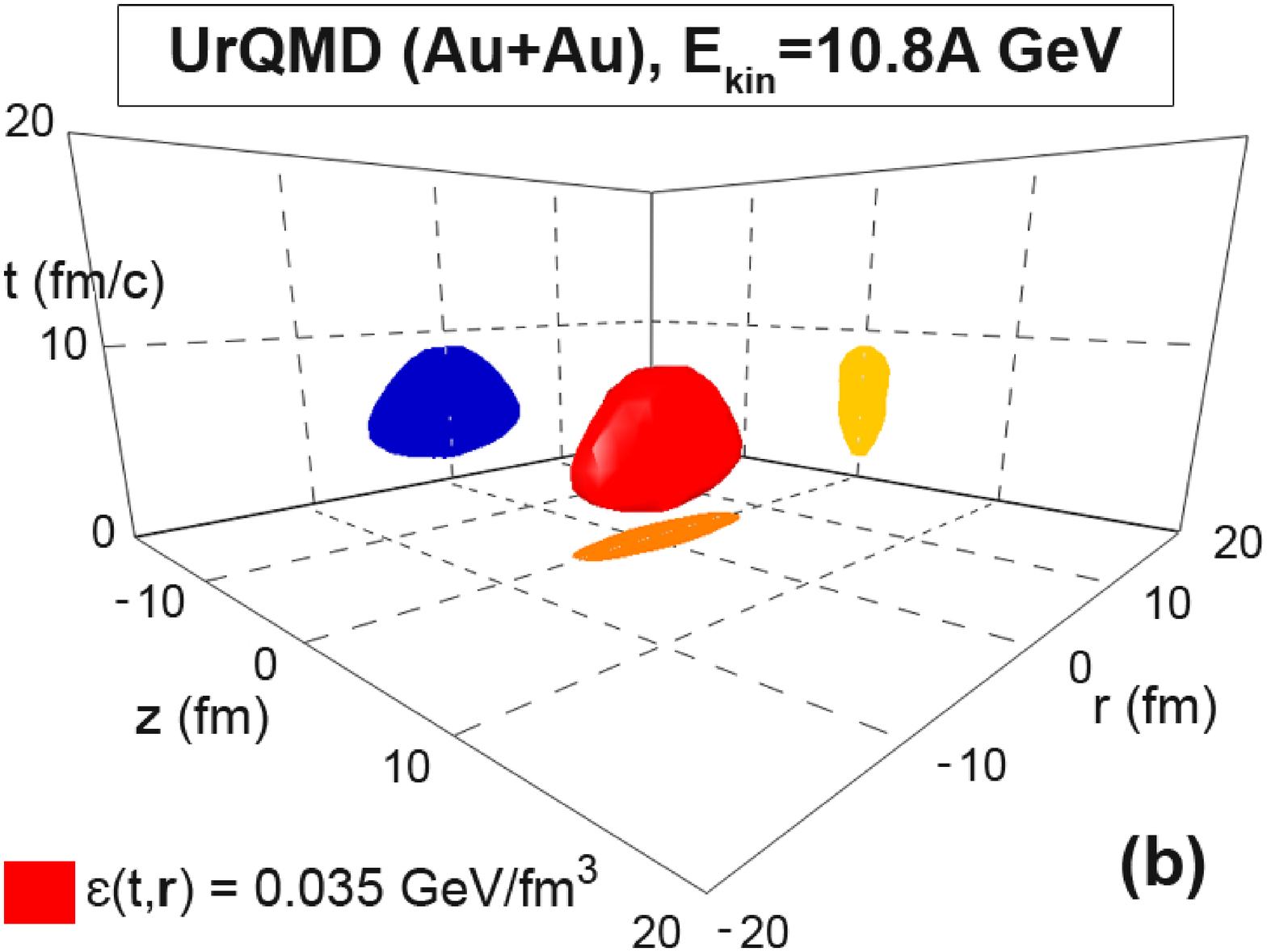}
\end{minipage}
\begin{minipage}{.6\textwidth}
\centering
\includegraphics[width=\textwidth]{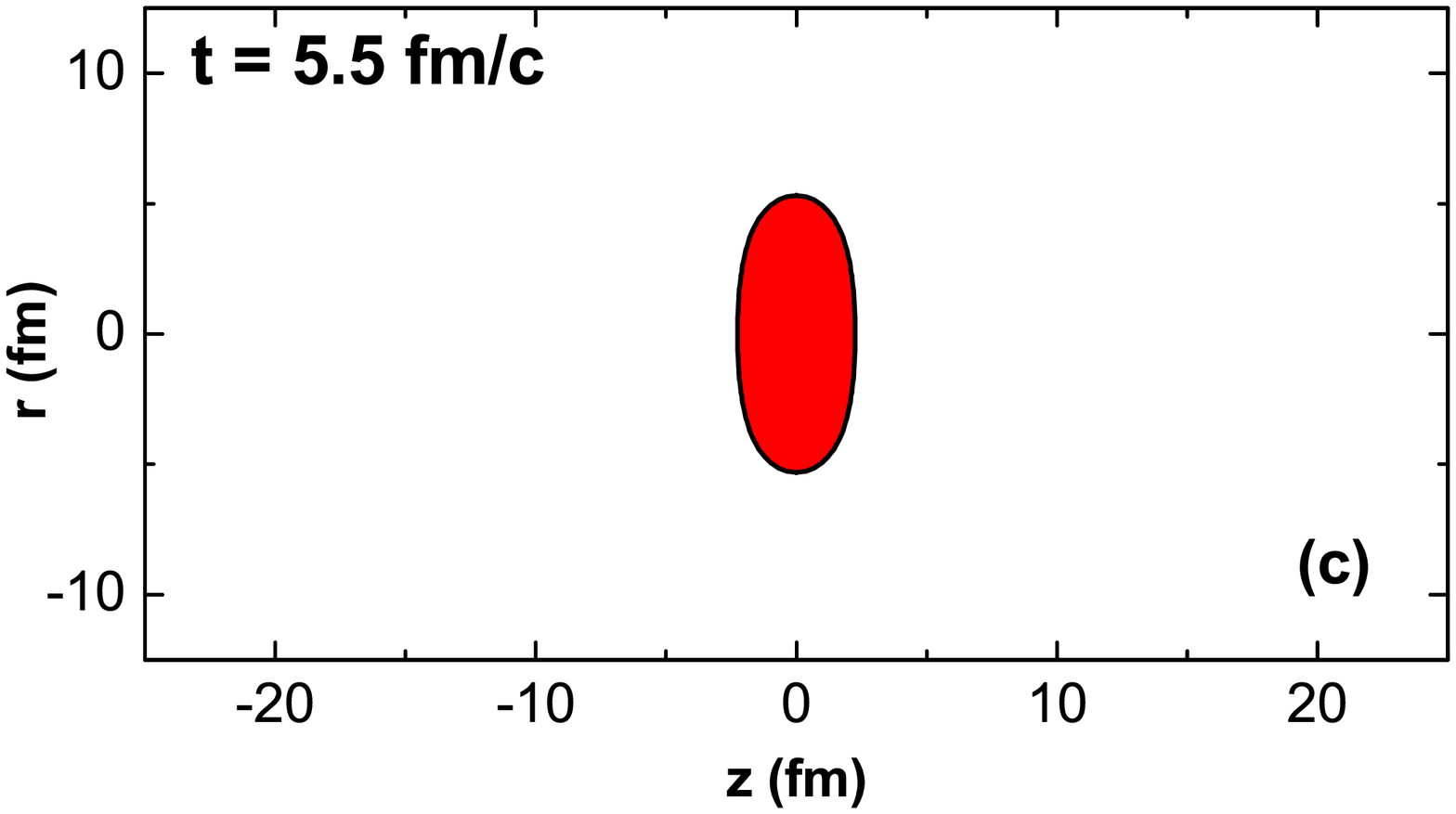}
\end{minipage}
\centering
\vspace{-0.3cm}
\caption{(Color online)
(a) Three-dimensional hypersurface of constant
invariant particle density of negative pions for AGS conditions
(E$_{\rm kin} = 10.8A$~GeV).
(b) The hypersurface at constant invariant energy density.
(c) The same hypersurface in $z$-$r$ coordinates at 5.5 fm/c.}
\label{fig:Hyp-AGS}
\end{figure}
divide it into cells with the side length of 1 fm.
For each cell the $\pi^-$ density, $n_{\rm lab} (x)$, and the average
velocity, $\bs v_{_E} (x)$ of pions in that cell are calculated as a result of
averaging over 1000 collision events.
Then the particle four-flow $N^{\mu}(x)$ for this cell is calculated by using
relation \eqref{eq:nmu}, which allows us to determine the Eckart
four-velocity $u^{\mu} (x)$, see eq. \eqref{eq:umu}.
Relation \eqref{eq:ninv} then allows to determine the $\pi^-$ invariant density
for each cell.

In order to determine the invariant energy density in a cell, the
energy-momentum tensor, $T^{\mu \nu} (x)$, is calculated in UrQMD by averaging
data over 1000 events according
to eq. \eqref{eq:Tmunu} for each cell.
The invariant pion energy density is then determined by eq. \eqref{eq:edens}.

\section{Results and discussion}

In order to determine the space-time characteristics of the fireball and FO
process it is useful to analyze the hypersurfaces
of constant invariant $\pi^-$ particle and energy densities.
Because of the symmetry of central collisions
the particle and energy densities do not depend on
azimuthal angle $\varphi$ in the $x$-$y$ plane
when transforming to cylindrical coordinates, that is
$n(t,x,y,z) = n(t,r,z)$ and $\epsilon(t,x,y,z) = \epsilon(t,r,z)$.
In this case it is possible to visualize the constant density hypersurfaces
in coordinates $(t,r,z)$, where $r=\pm \sqrt{x^2+y^2}$.

\begin{figure}[ht]
\begin{minipage}{.46\textwidth}
\centering
\includegraphics[width=\textwidth]{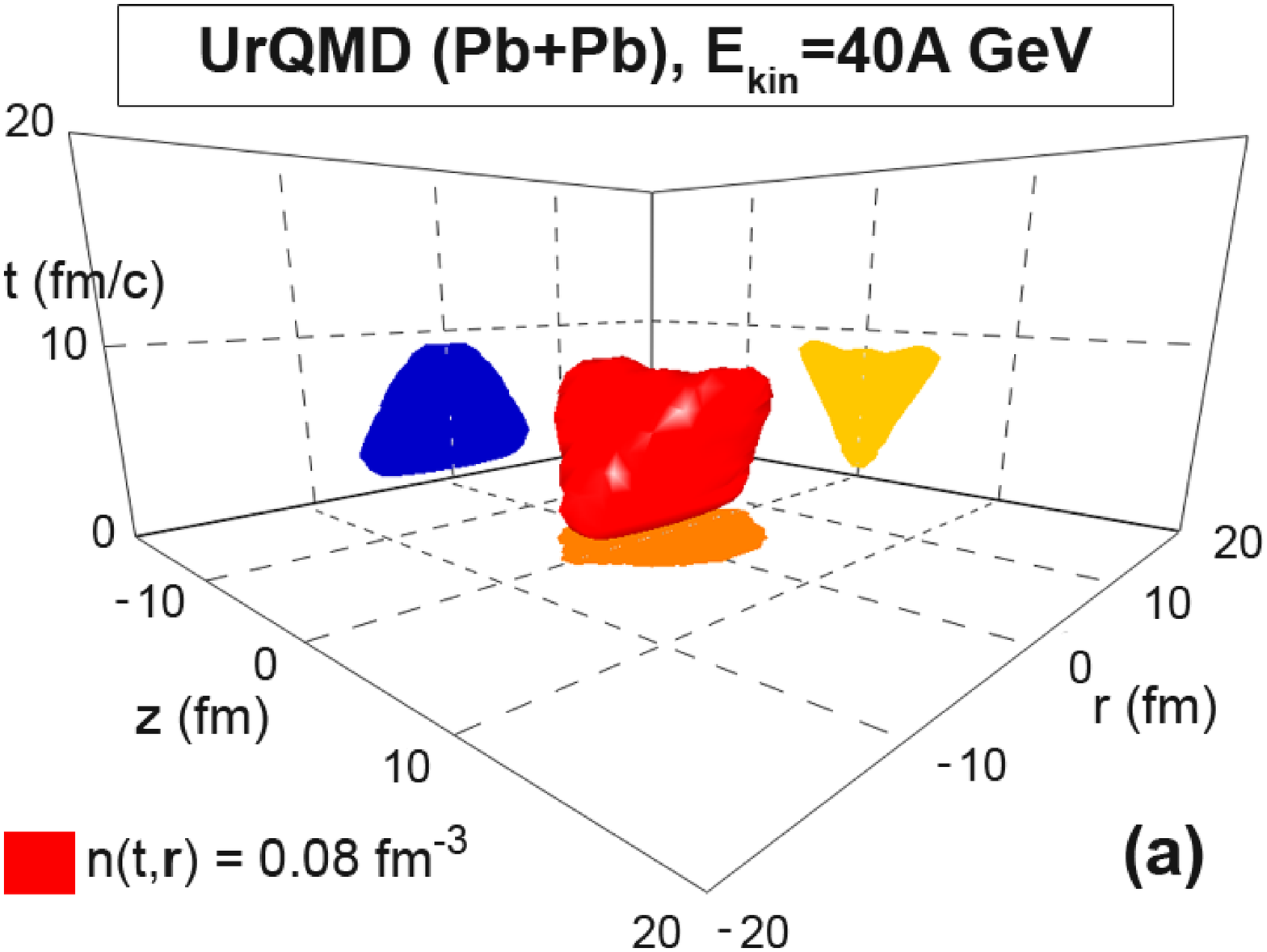}
\end{minipage}
\begin{minipage}{.46\textwidth}
\centering
\includegraphics[width=\textwidth]{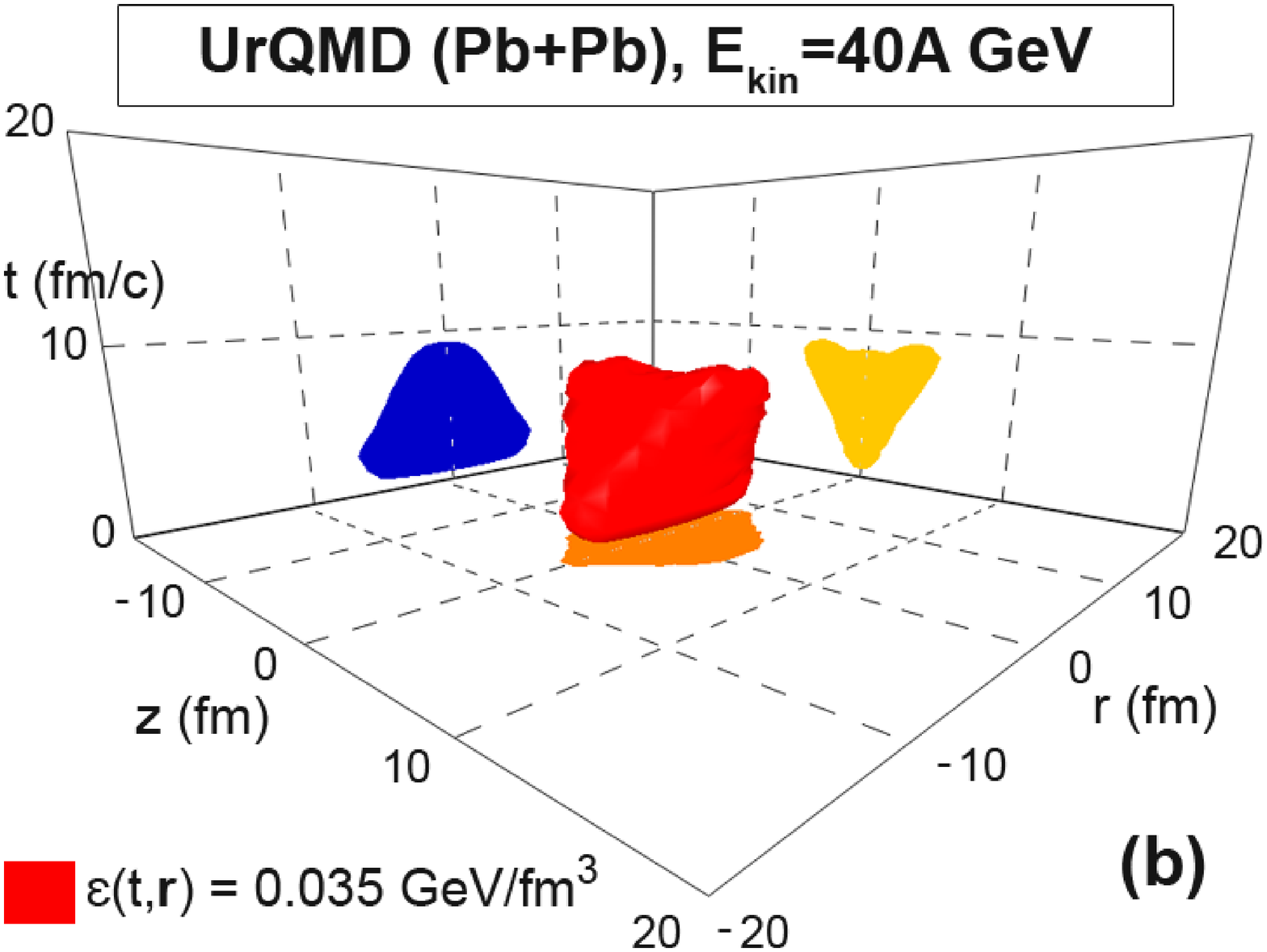}
\end{minipage}
\centering
\vspace{-0.3cm}
\caption{(Color online) Same as Fig. \ref{fig:Hyp-AGS} but for SPS conditions ($E_{\rm kin} = 40A$~GeV).}
\label{fig:Hyp-lowSPS}
\end{figure}

\begin{figure}[h]
\begin{minipage}{.46\textwidth}
\centering
\includegraphics[width=\textwidth]{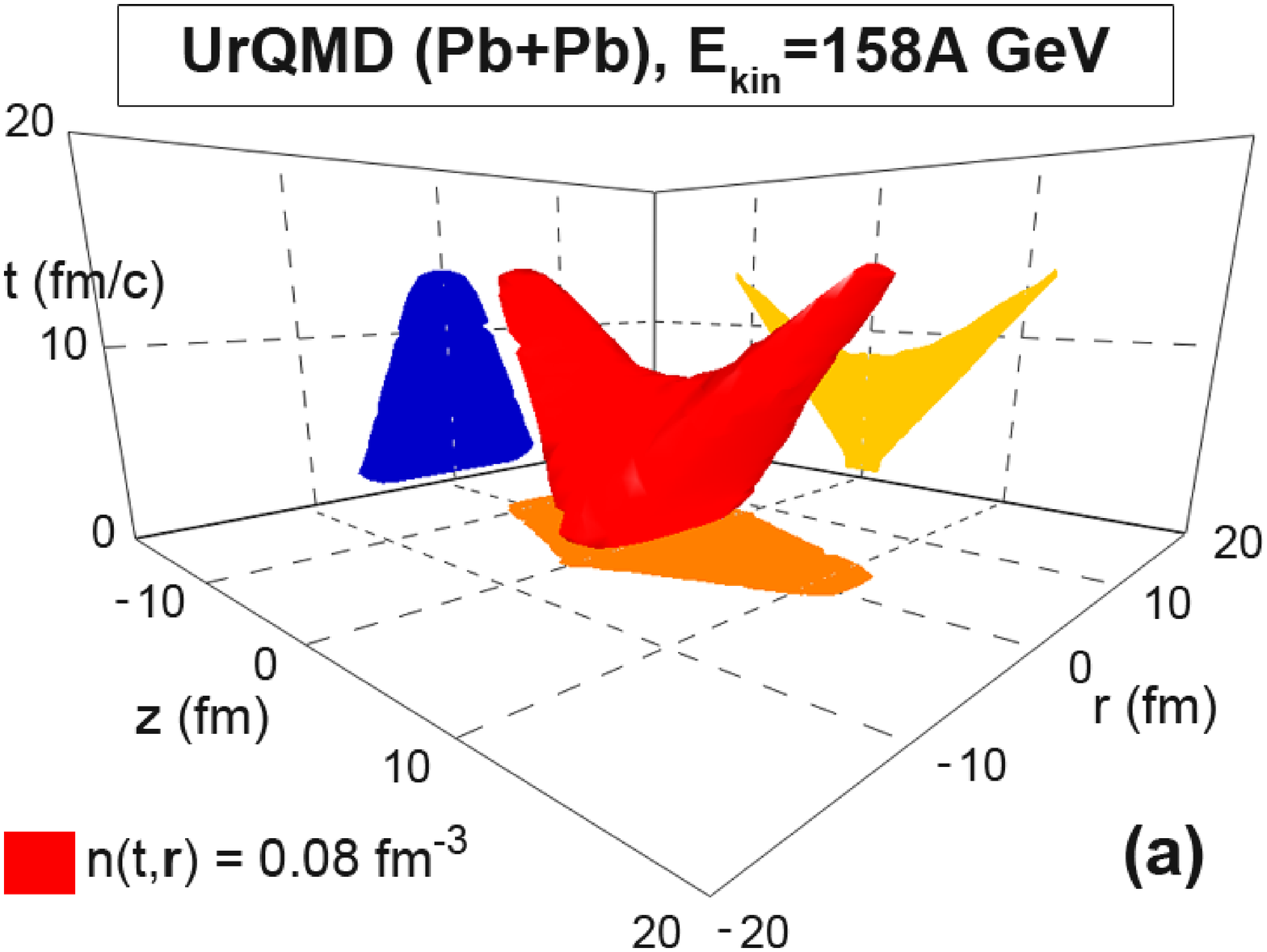}
\end{minipage}
\begin{minipage}{.46\textwidth}
\centering
\includegraphics[width=\textwidth]{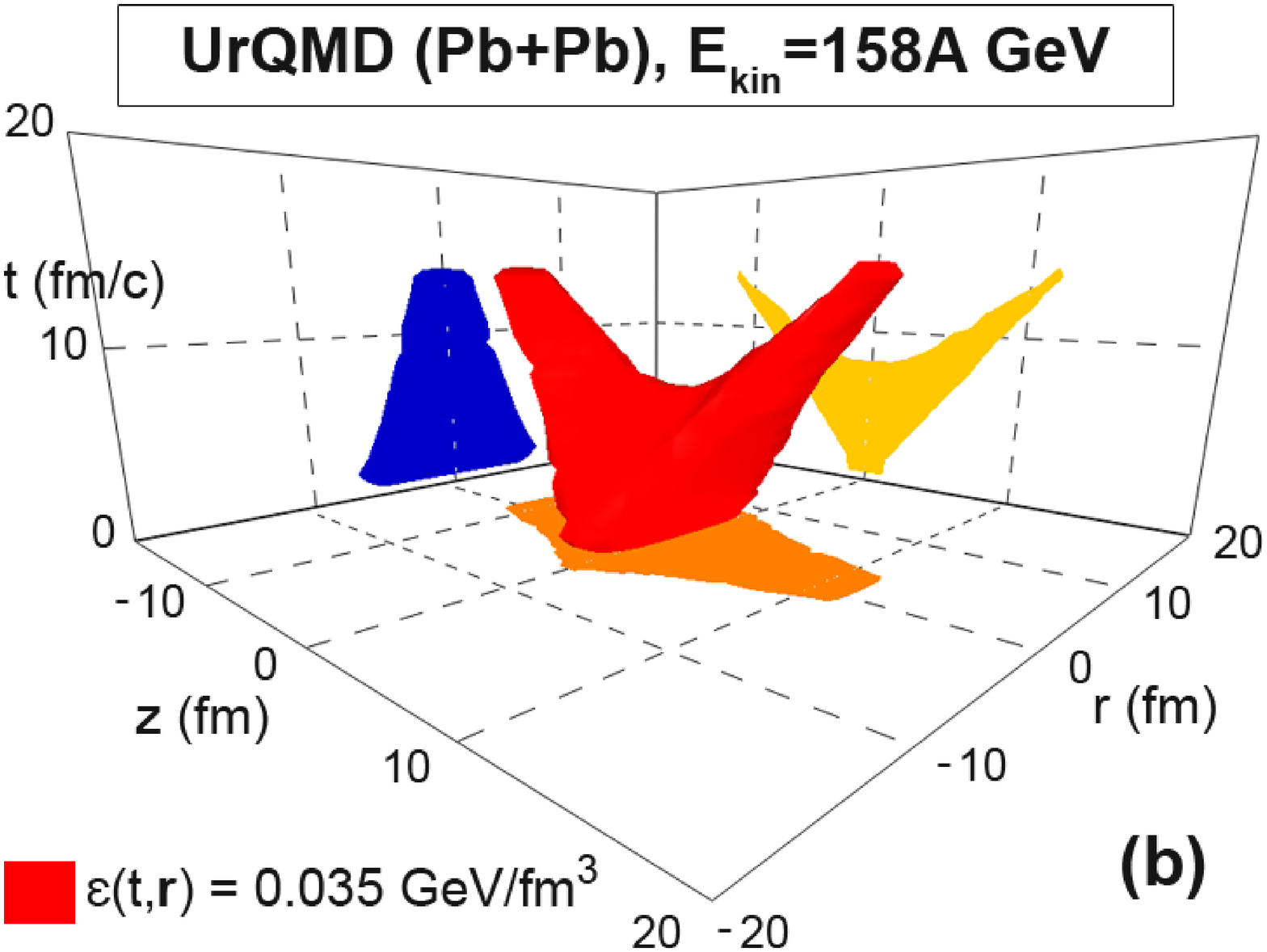}
\end{minipage}
\begin{minipage}{.6\textwidth}
\centering
\includegraphics[width=\textwidth]{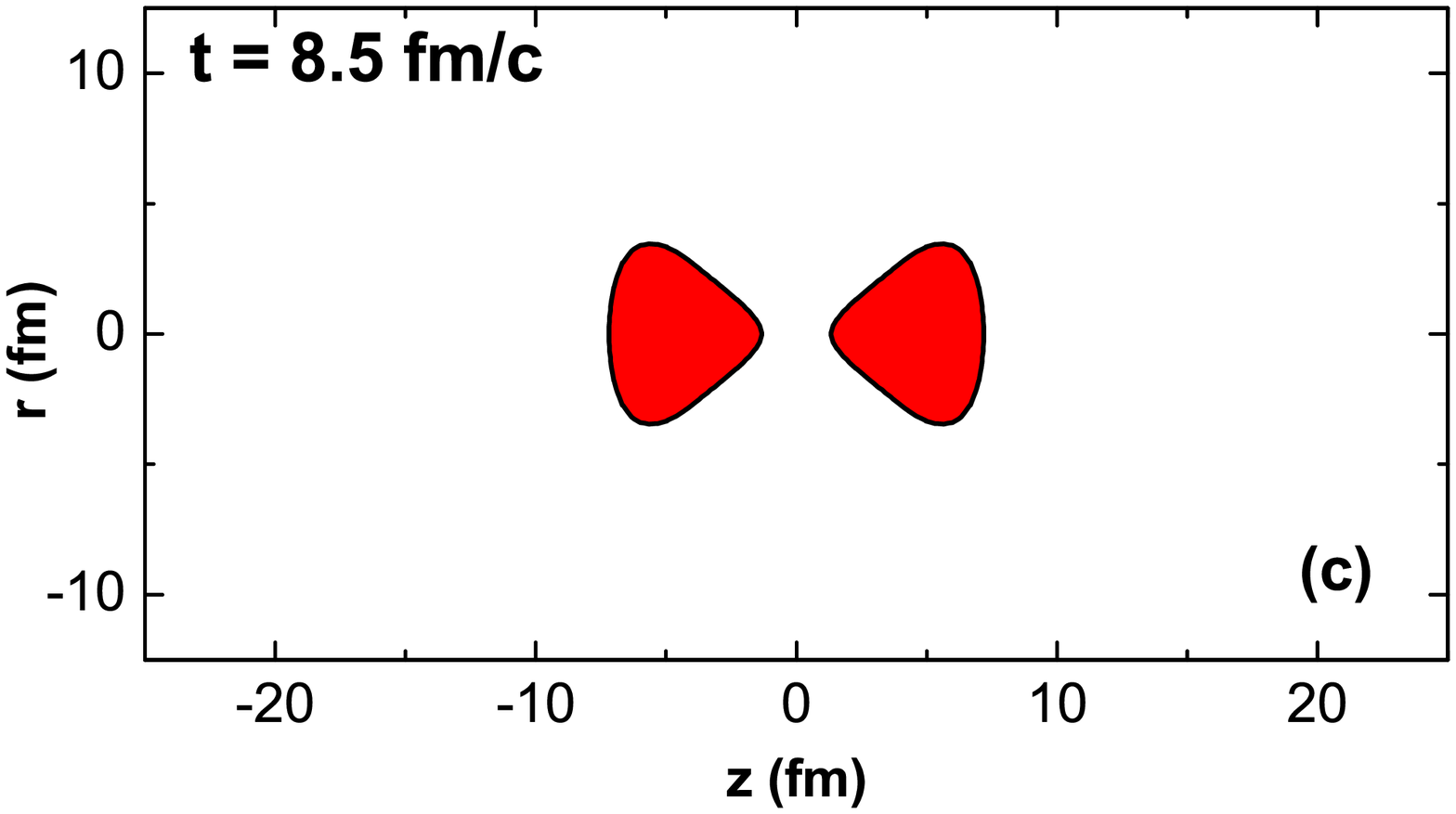}
\end{minipage}
\centering
\vspace{-0.3cm}
\caption{(Color online) Same as Fig.~\ref{fig:Hyp-AGS} but for SPS conditions ($E_{\rm kin} = 158A$~GeV).}
\label{fig:Hyp-SPS}
\end{figure}

\begin{figure}
\begin{minipage}{.46\textwidth}
\centering
\includegraphics[width=\textwidth]{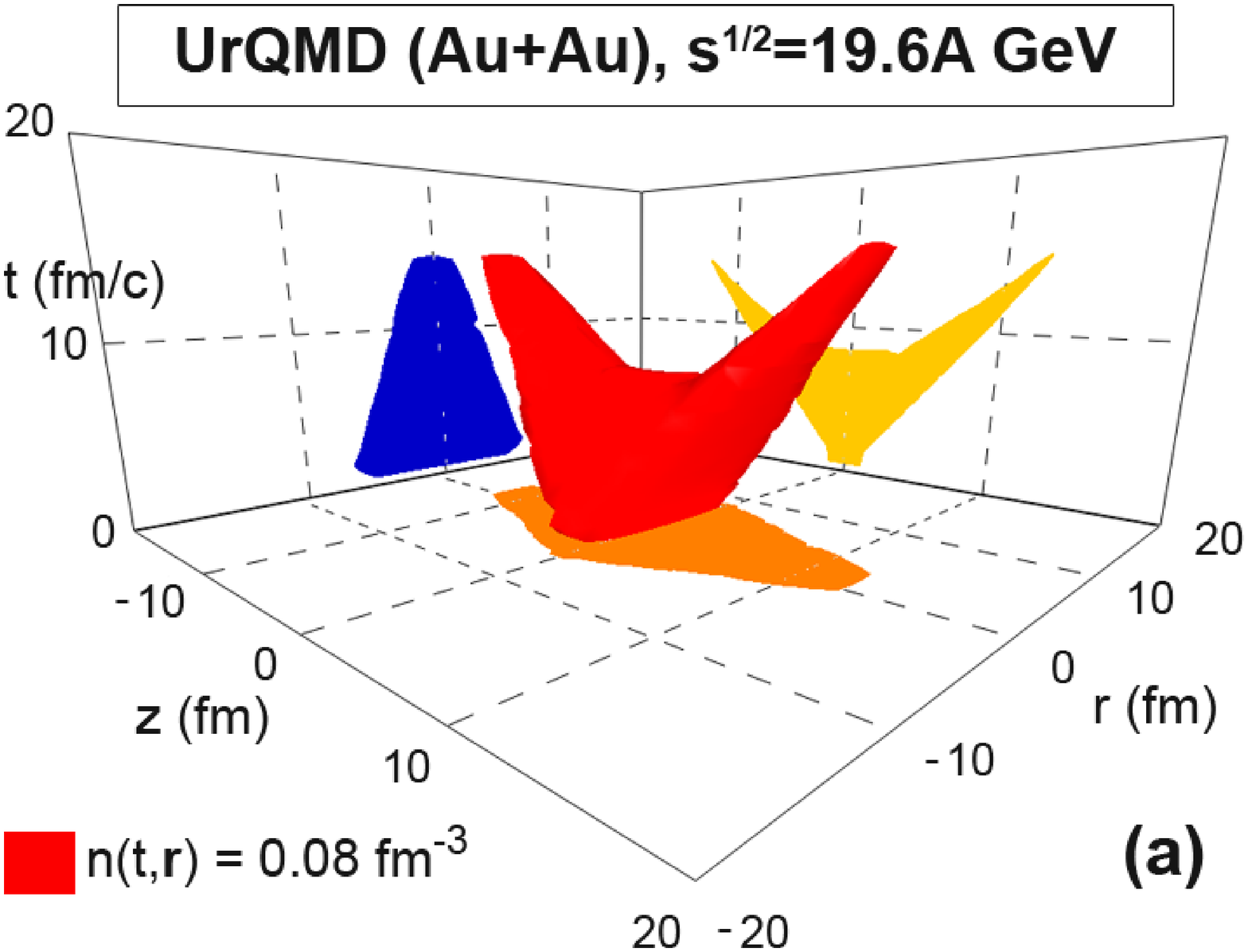}
\end{minipage}
\begin{minipage}{.46\textwidth}
\centering
\includegraphics[width=\textwidth]{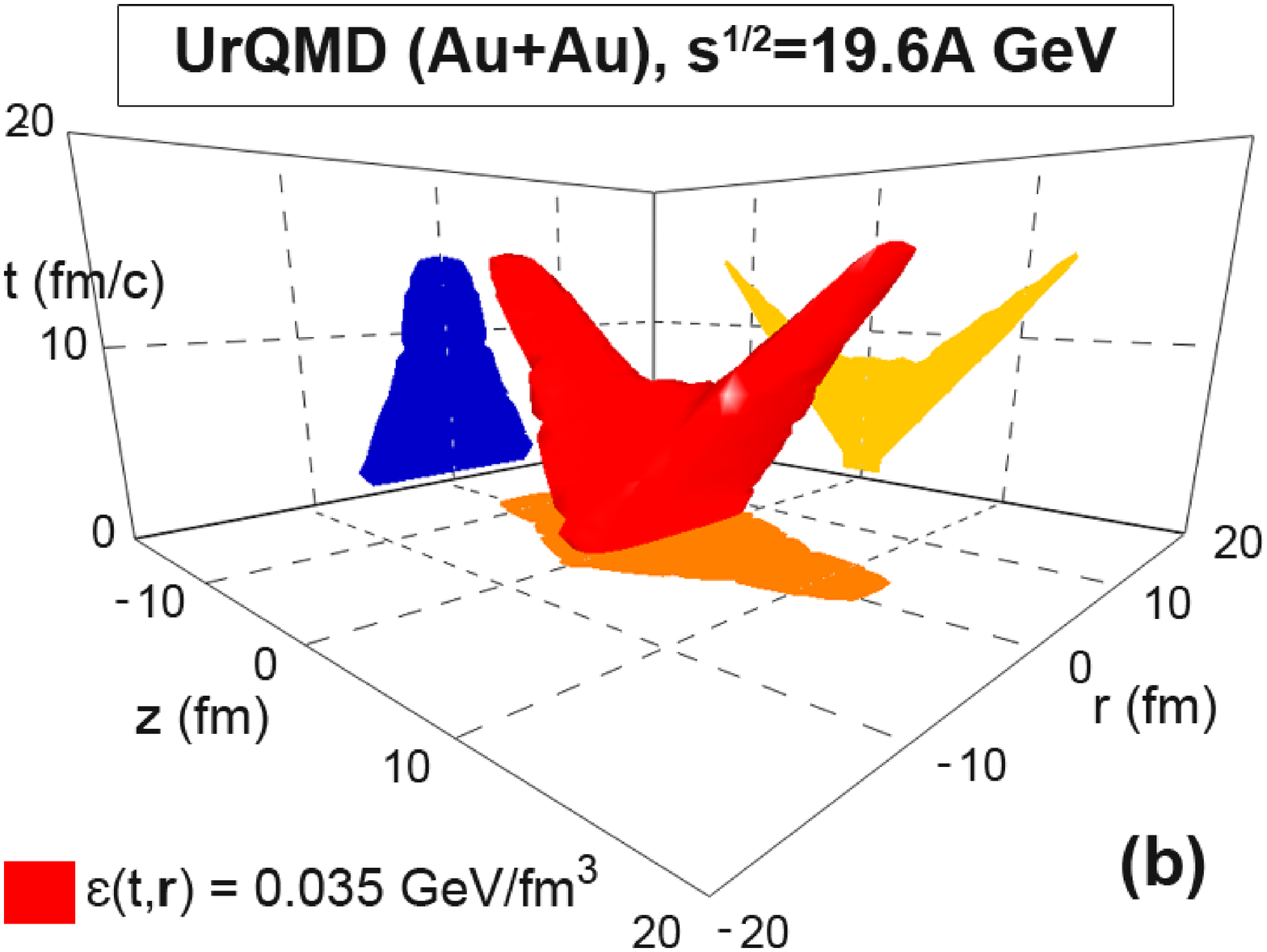}
\end{minipage}
\centering
\vspace{-0.3cm}
\caption{(Color online) Same as Fig.~\ref{fig:Hyp-AGS} but for RHIC conditions
($\sqrt{s_{AA}} = 19.6A$~GeV).}
\label{fig:Hyp-RHIC-1}
\end{figure}

\begin{figure}
\begin{minipage}{.46\textwidth}
\centering
\includegraphics[width=\textwidth]{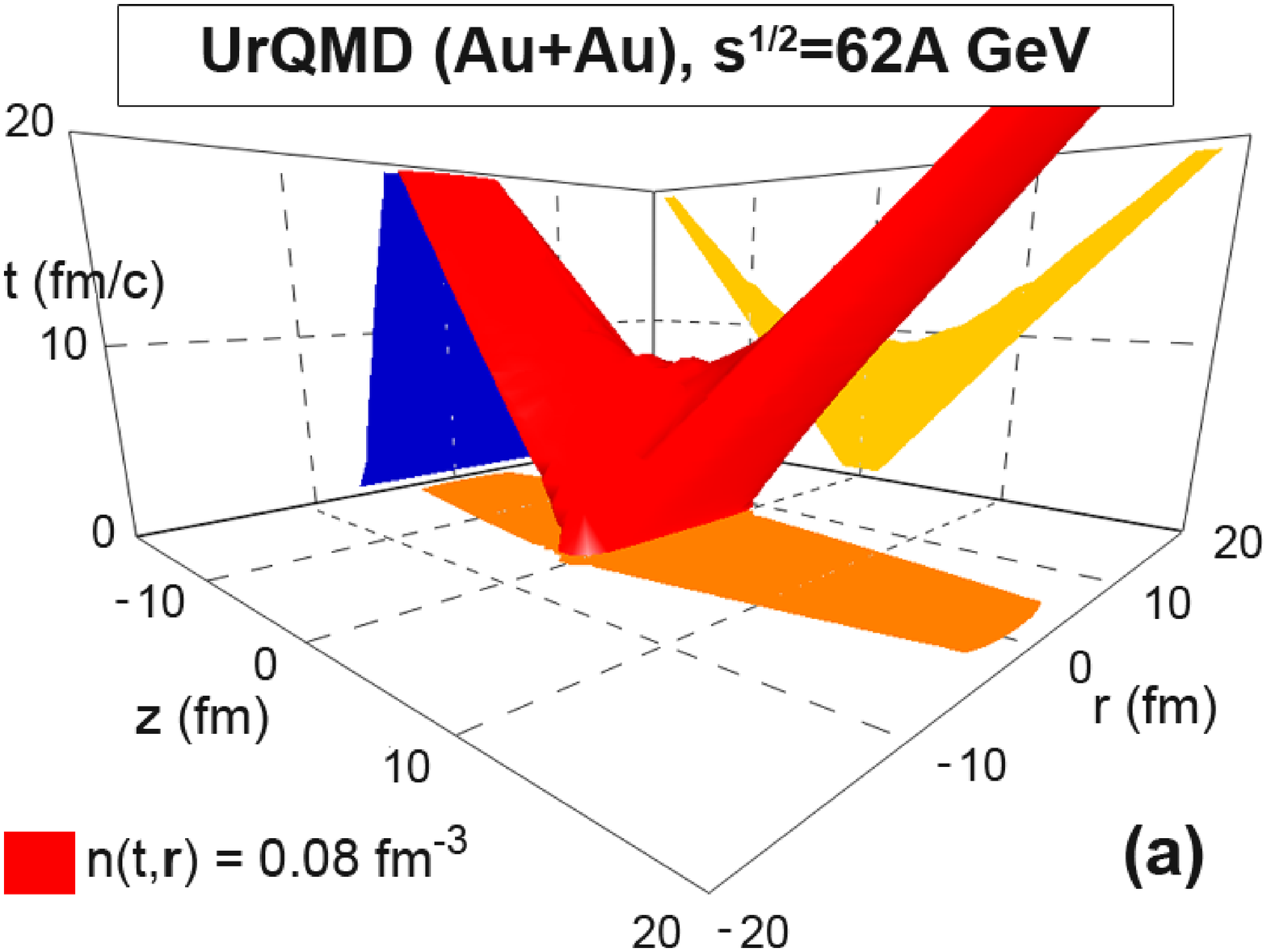}
\end{minipage}
\begin{minipage}{.46\textwidth}
\centering
\includegraphics[width=\textwidth]{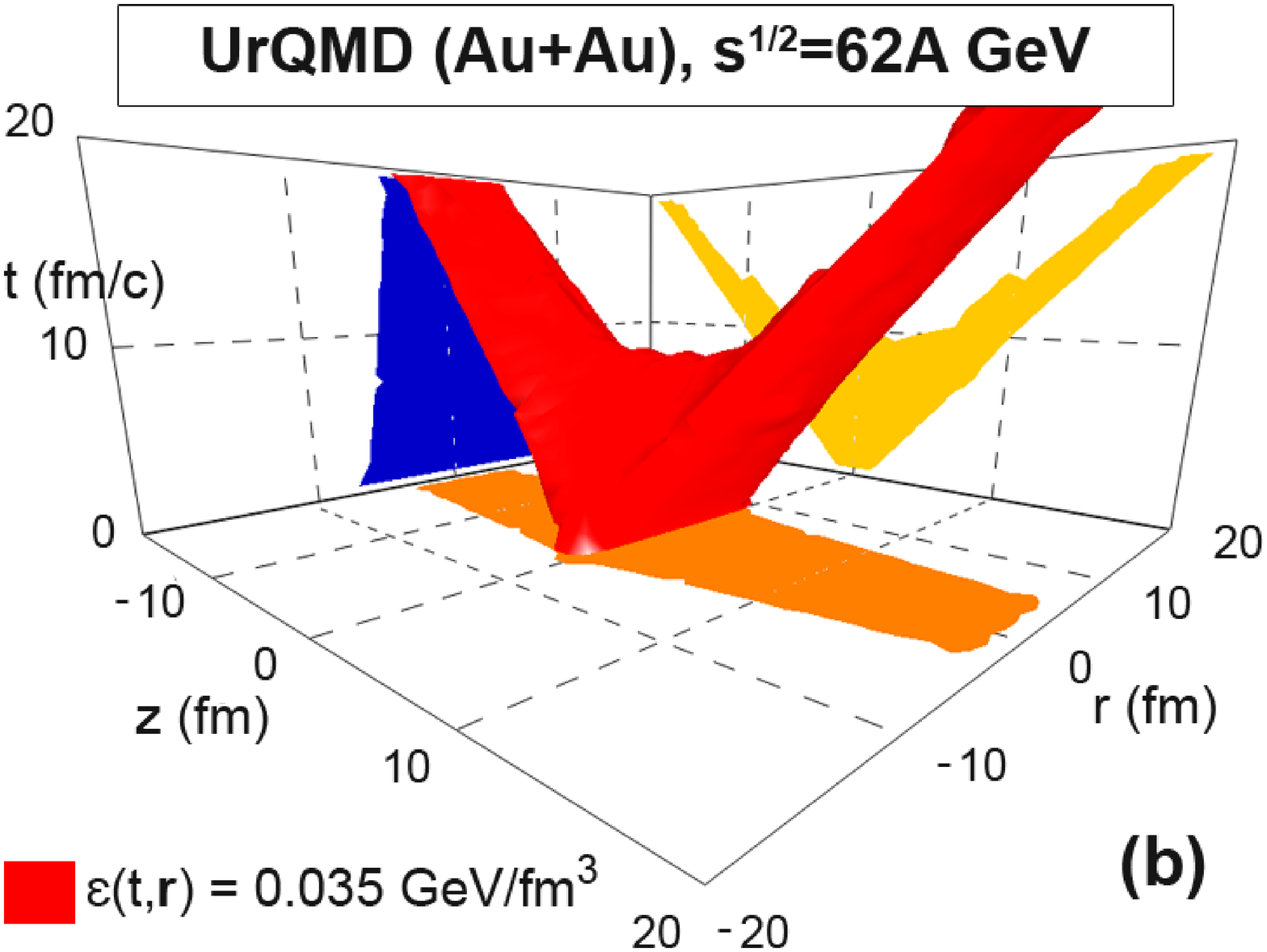}
\end{minipage}
\centering
\vspace{-0.3cm}
\caption{(Color online) Same as Fig.~\ref{fig:Hyp-RHIC-1} but for
$\sqrt{s_{AA}} = 62A$~GeV.}
\label{fig:Hyp-RHIC-2}
\end{figure}

Calculation results for $n(x) = n_c = 0.08$~/fm$^3$ and
$\epsilon(x) = \epsilon_c = 0.035$~GeV/fm$^3$ are
depicted in Figs. \ref{fig:Hyp-AGS}-\ref{fig:Hyp-RHIC-2}
for central collisions with energies ranging from AGS to RHIC.
It might be mentioned that the UrQMDv2.3 model is not known to yield
accurate results at RHIC energies where extremely hot and dense nuclear matter
is created in the early stage of collision. However, it is still useful
to analyze the FO hypersurface within this model and
to obtain at least qualitative space-time features of
the FO process and to compare them with corresponding features at lower energies.

\begin{figure}
\vspace{3.5mm}
\begin{minipage}{.46\textwidth}
\centering
\includegraphics[width=\textwidth]{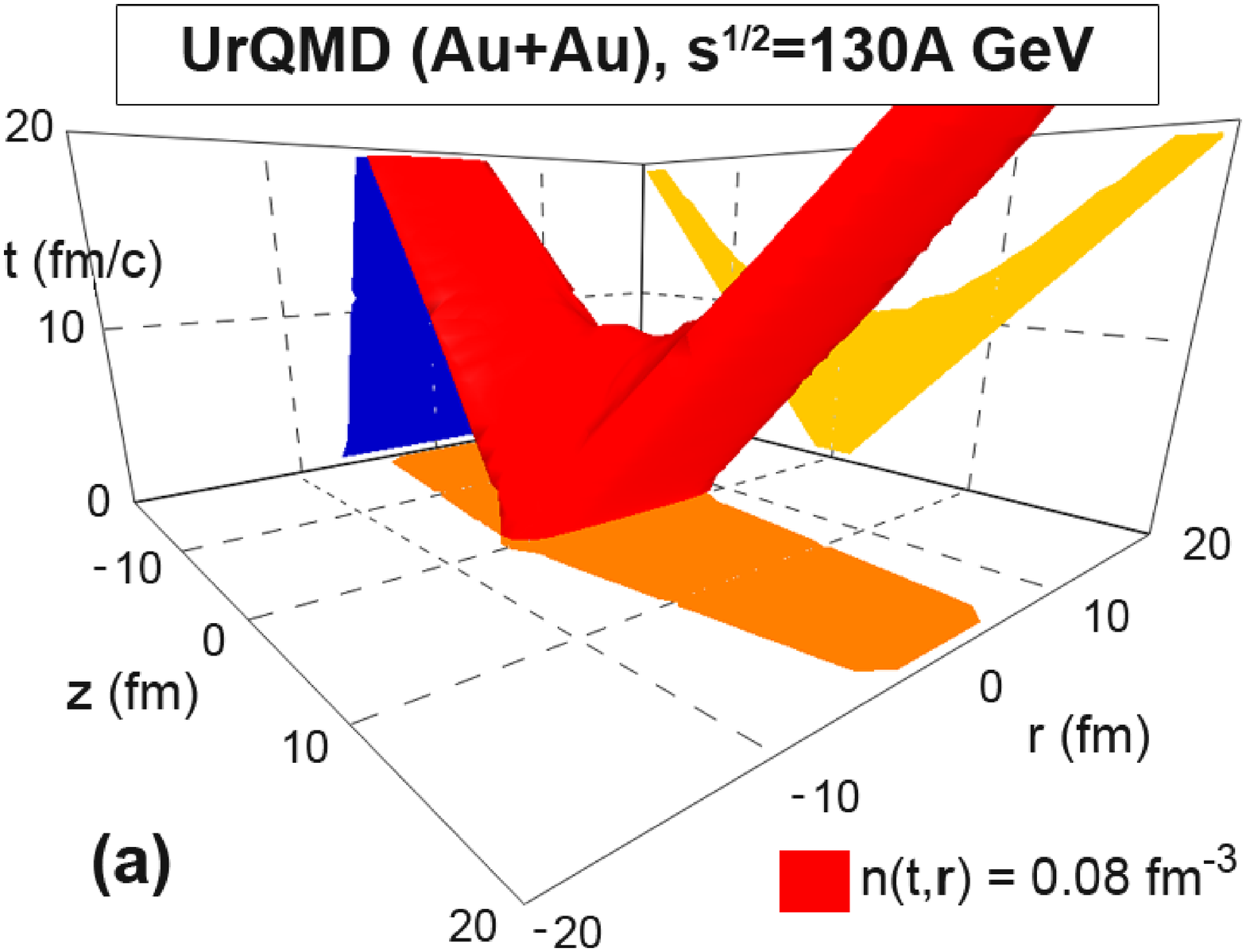}
\end{minipage}
\begin{minipage}{.46\textwidth}
\centering
\includegraphics[width=\textwidth]{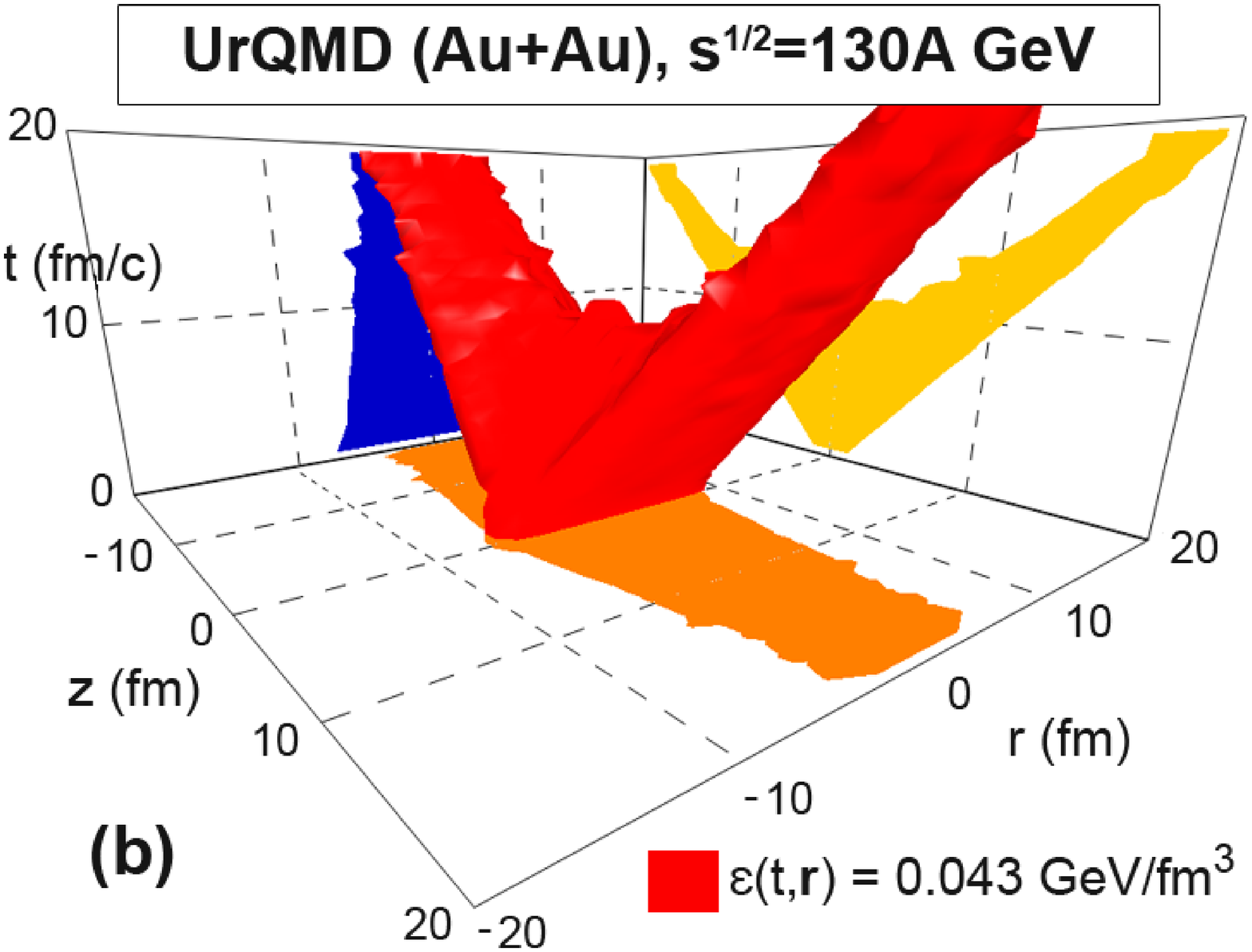}
\end{minipage}
\centering
\vspace{-0.3cm}
\caption{(Color online) Same as Fig.~\ref{fig:Hyp-RHIC-1} but for
$\sqrt{s_{AA}} = 130A$~GeV.}
\label{fig:Hyp-RHIC-3}
\end{figure}

It can be seen that the hypersurfaces, corresponding to constant particle
and energy density virtually coincide.
In fact, the value of $\epsilon_c = 0.035$~GeV/fm$^3$ was
specially chosen to show (to prove) that both FO hypersurfaces coincide with one
another when they are determined by means of equations $n(x) = n_c$ and
$\epsilon(x) = \epsilon_c$ provided by specific correspondence of
$n_c$ to $\epsilon_c$ (see Figs. \ref{fig:Hyp-AGS}-\ref{fig:Hyp-RHIC-2}).
In case of a higher RHIC energy ($\sqrt{s_{AA}}=130A$~GeV) the
corresponding value of $\epsilon_c$ is higher and is equal to $0.043$~GeV/fm$^3$
(see Fig. \ref{fig:Hyp-RHIC-3}).

\bigskip

\bigskip

Regarding these hypersurfaces as sharp FO hypersurfaces we can
determine temperature $T_{\rm f}$ at FO by using relation
\begin{equation}
\frac{\epsilon_c}{n_c}\ =\ 3T_{\rm f}\, +\,
m_\pi \frac{K_1(m_\pi/T_{\rm f})}{K_2(m_\pi/T_{\rm f})} \,,
\end{equation}
assuming that at the FO hypersurface we are still having an ideal,
dilute pion gas described with the Boltzmann single-particle distribution,
$f_{_{\rm B}}(\bs p) \propto \exp{(-\sqrt{m_{\pi}^2+\bs p^2}/T_{\rm f})}$, in
the rest frame of the element of FO hypersurface.
Here $K_1$ and $K_2$ are the modified Bessel functions of the second kind.

Solving the above equation yields $T_{\rm f} = 128$~MeV for AGS, SPS and lower
RHIC energies.
For higher RHIC energy of $\sqrt{s_{AA}}=130A$~GeV the temperature
rises to 164~MeV.
As in our calculation we assume the validity of the UrQMD dynamics both before
and after FO, we have only one FO temperature $T_f$.
If we would replace the pre-FO stage by fluid dynamics then the pre-FO
temperature may be different, while the conserved currents across the front
must be conserved.

It is worth mentioning that the same value of temperature, calculated
by the above formula, also appears in other space-time regions.
That is, there are several hypersurfaces corresponding to the same value of
temperature so it is only possible to define temperature at FO, but it is
not possible to define a single, unique, connected FO hypersurface  of
constant temperature. Especially at higher energies, the constant temperature
hypersurface fragments to unconnected pieces.

We can define the lifetime $\tau$ of the pion system (pion fireball) confined
by the FO hypersurface as a maximum time moment of the projection of FO
hypersurface on $t-z$ plane.
It is seen that at AGS energies (see Fig. \ref{fig:Hyp-AGS} and Table
\ref{tab:space-time-rad}) the fireball has a lifetime of about
$\tau=9$~fm/$c$, and maximum transverse and longitudinal radii of 6 and 2 fm
respectively.
The exact values of these parameters would change for different choice of $n_c$
but it can be seen that for this energies the longitudinal expansion of pion
fireball is rather small.
One qualitative feature of pion fireball at AGS energies is that it always
stays spatially as a single freezing out object during the whole time
evolution.

\begin{table}
 \caption{{\bf Space-time evolution parameters.}} 
 \centering                                                 
 \begin{tabular}[c]{c|c|c|c|c|c}                              
  $E_{\rm kin}$ & $\sqrt{s_{AA}}$ & $A+A$ &$\tau$  & $R_{\perp}$ & $R_{\parallel}$\\
   (A GeV)&  (A GeV)&  &  (fm$/c$)&  (fm)&  (fm)\\
 \hline\hline                                               
 10.8  &  4.88 & $Au+Au$  & 9 & 6  &  2   \\
 \hline
 20.0 & 6.41   & $Pb+Pb$ & 9 & 6 &  3    \\
 40.0 & 8.86   &         & 8.75 & 6.5 & 5    \\
 80.0 & 12.39  &         & 8.75 & 6.5 & 7    \\
 158.0 & 17.32 &         & 8.5 & 6.5 & 7.5  \\
 \hline
 202.9 & 19.6   & $Au+Au$ & 8.25 & 6.5 &  7.5    \\
 2047.0 & 62.0   &         & 8.75 & 6.5 & 8.75    \\
 9007.0 & 130.0  &         & 10 & 6.5 & 10    \\
 \hline
 \end{tabular}
 \label{tab:space-time-rad}
\end{table}

The situation changes at SPS energies (see Figs.~\ref{fig:Hyp-lowSPS} and
Table \ref{tab:space-time-rad}).
It is seen that after some time $t_{\rm d} \approx 9$~fm/$c$ the pion fireball
spatially breaks up into two parts (see the projection of FO
hypersurface onto $t-z$ plane).
That is, $t_{\rm d}$ is actually a fireball division time.

Hence, at the time $t=t_{\rm d}$ the fireball is separating into two
pieces (drops), which lifetimes are counting starting from this moment.
By considering the fireball just as a whole, connected body in the
4-dimensional space-time (i.e. a single piece of nuclear matter) we
can generalize the definition of the fireball lifetime $\tau$.
So, we define the lifetime of the fireball as that time span when
fireball exists just as one single piece of exited nuclear matter, i.e.
$\tau \equiv t_{\rm d}$.
After division of the fireball there are two drops of matter which move in
opposite directions from one another with velocities
approaching the speed of light with increase of collision energy
(see Figs.~\ref{fig:Hyp-lowSPS}--\ref{fig:Hyp-RHIC-3}, especially
upper and lower panels of Fig.~\ref{fig:Hyp-SPS}).

With the increase of energy the longitudinal radius increases while the
fireball division time $t_{\rm d}$ stays about the same.
Since the velocity of pions cannot exceed the speed of light and the fireball
formation time is very small compared to the fireball lifetime, the longitudinal
radius is bounded from above by $\tau$.
It reaches it's maximum value at SPS energies of about 7~fm,
which is consistent with HBT interferometry measurements \cite{Alt2008radii}.

At higher energies available at RHIC (Figs.~\ref{fig:Hyp-RHIC-1} and
Table~\ref{tab:space-time-rad}) the picture is similar to SPS energies:
the fireball expands as a whole until about $t_{\rm d} \simeq 8.25-10$~fm/$c$
and then breaks up into two parts.
The values of $R_{||}$ are consistent with HBT radii
\cite{Adler2001radii,Chen2009}.

\begin{table*}
 \caption{{\bf Fireball lifetime for different values of $n_c$}} 
 \centering                                                 
 \begin{tabular}[c]{c|c|c|c|c|c}                              
  $E_{\rm kin}$ (A GeV)& $\sqrt{s_{AA}}$ (A GeV)& $A+A$ &
  \multicolumn{3}{ c }{Fireball lifetime (fm$/c$)}\\
  \hline
   & & & $n_c = 0.04$~fm$^{-3}$ & $n_c = 0.08$~fm$^{-3}$ & $n_c = 0.12$~fm$^{-3}$\\
 \hline\hline                                               
 10.8  &  4.88 & $Au+Au$  & 12 & 9  &  7  \\
 \hline
 20.0 & 6.41   & $Pb+Pb$ & 12 & 9 &  7   \\
 40.0 & 8.86   &         & 12 & 8.75 &  7   \\
 80.0 & 12.39  &         & 12.75 & 8.75 & 6.5  \\
 158.0 & 17.32 &         & 13.5 & 8.5 & 6.25 \\
 \hline
 202.9 & 19.6   & $Au+Au$ & 13 & 8.25 &  6.25   \\
 2047.0 & 62.0   &         & 13 & 8.75 & 6.75   \\
 9007.0 & 130.0  &         & 14 & 10 & 7.5   \\
 \hline
 \end{tabular}
 \label{tab:tfd-nc}
\end{table*}
What indicates that the fireball division times, $t_{\rm d}$, and thus the
fireball lifetimes, $\tau$, change very weakly with collision energy
(see Table~\ref{tab:space-time-rad}), they vary in the range
$\tau \simeq 8.25-10.0$~fm/$c$.

The fireball space-time structure depends on the chosen value of
$n_c$ (and/or, respectively on the value of $\epsilon_c$).
The value of $n_c = 0.08$~fm$^{-3}$ was chosen so that maximum fireball size
at SPS and RHIC energies corresponds to known interferometry data
$R \sim 6-8$~fm.
It can be argued what is the exact value of $n_c$ that corresponds to the
sharp kinetic (or chemical) FO hypersurface; so, it is useful
to investigate the structure of the fireball at different values of $n_c$.

One feature of the fireball structure for $n_c = 0.08$~fm$^{-3}$ is that the
fireball lifetime, $\tau$ depends weakly on collision energy
(see Table~\ref{tab:space-time-rad}).
The values of $\tau$ for different values of $n_c$ and for different collisions
energies are presented in Table \ref{tab:tfd-nc}.
It is seen that $\tau$ depends on the $n_c$ value (with the increase of $n_c$
the value of $\tau$ decreases), but for each $n_c$ it depends very weakly on
collision energy: for $n_c = 0.04$~fm$^{-3}$ it is in the range
$\tau \simeq 12-14$~fm/$c$ and for $n_c = 0.12$~fm$^{-3}$ lifetime is in the range
$\tau \simeq 6.25-7.5$~fm/$c$.
Thus, it can be stated that {\it the fireball lifetime is approximately
invariant of the  collision energy}.

\begin{figure*}
\centering
\includegraphics[width=.65\textwidth]{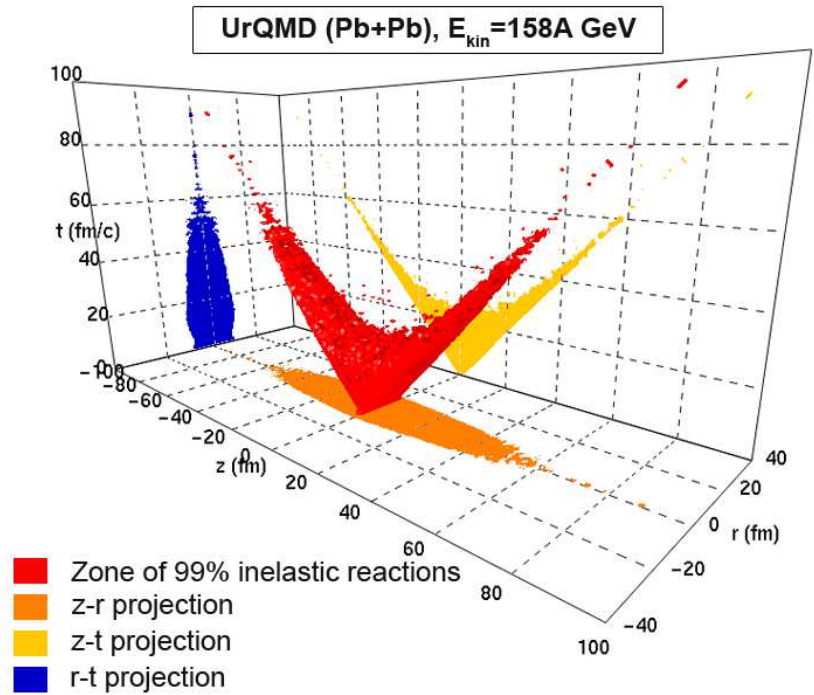}
\centering
\vspace{-0.3cm}
\caption{(Color online) The three-dimensional reaction zone, which determines
the space-time region where 99\% of all inelastic hadronic reactions
for SPS conditions ($E_{\rm kin} = 158$A GeV) take place.}
\label{fig:ZoR-3D-Pb158158}
\end{figure*}
\begin{table}[!h]
 \caption{{\bf  Division times of the reaction zone
 (different fractions of the contained inelastic reactions) }} 
 \centering                                                 
 \begin{tabular}[c]{c|c|c|c|c}                              
  $E_{\rm kin}$& $\sqrt{s_{AA}}$ & $A+A$ &   $t_d~(99\%)$ &   $t_d~(80\%)$\\
 (A~GeV) &  (A~GeV) &  &  (fm/$c$) & (fm/$c$)\\
 \hline\hline                                               
 10.8  &  4.88 & $Au+Au$ & 17.5 & 8.5\\
 \hline
 20.0 & 6.41   & $Pb+Pb$ & 17 & 7.5\\
 40.0 & 8.86   &         & 17.5 & 7 \\
 80.0 & 12.39  &         & 18 & 7\\
 158.0 & 17.32 &         & 19.5 & 7\\
 \hline
 \end{tabular}
 \label{tab:trz}
\end{table}
The hypersurfaces of constant $\pi^-$ particle and energy density can be
compared to hadron reaction zones.
Reaction zone is defined as the space-time region where a certain fraction of
reactions of certain type took place (see Ref.~\cite{Anchishkin2010}).
The reaction zone, which contains 99\% of all inelastic hadronic reactions
(see. Fig.~\ref{fig:ZoR-3D-Pb158158}) is related to $\pi^-$ FO process
(since inelastic reactions mostly involve pions) and includes the hypersurfaces
of constant $\pi^-$ particle and energy densities (which can be regarded as
sharp FO hypersurfaces).
This particular reaction zone is called hot fireball (see Ref.~\cite{Anchishkin2010}).
It is also seen that the reaction zone boundary (see Fig.~\ref{fig:ZoR-3D-Pb158158})
is similar to the hypersurfaces of constant particle density:
at SPS and RHIC energies the fireball expands as a whole for some time, $t$,
and then spatially breaks up into two parts, extending from each other in
the $\pm\, z$-directions.

Similarly to the case of constant particle and energy density
hypersurfaces, the reaction zone division time depends
weakly on collision energy (see Table~\ref{tab:trz}).
It is seen from Table~\ref{tab:trz} that the values of reaction zone division
time depend on the fraction of total inelastic hadronic reactions
which are contained in that reaction zone, for instance 80\% or 99\% of all
inelastic reactions.
It is similar to the dependence of fireball lifetime on the value of $n_c$
(see Table~\ref{tab:tfd-nc}).
Hot fireball (reaction zone containing 99\% of inelastic hadronic reactions)
division times are higher than corresponding values of lifetimes (division times) in
Table~\ref{tab:tfd-nc} regarding constant $\pi^-$ density hypersurfaces,
therefore it corresponds to a rather low $n_c$ for negative pions.
A different (lower) choice of percentage of inelastic reactions contained in
reaction zone, for instance 80\%, would correspond better to the constant $\pi^-$ density
hypersurfaces and their division times studied in this work
(see Table~\ref{tab:trz}, division times for reaction zone containing 80\%
of inelastic reactions).
From this we can conclude that the pion FO hypersurface, which is determined
in accordance with the critical density $n_c = 0.08$~fm$^{-3}$, contains the
space-time volume where approximately 80\% of all inelastic hadronic reactions
take place.

\begin{figure}
\begin{minipage}{.48\textwidth}
\centering
\includegraphics[width=\textwidth]{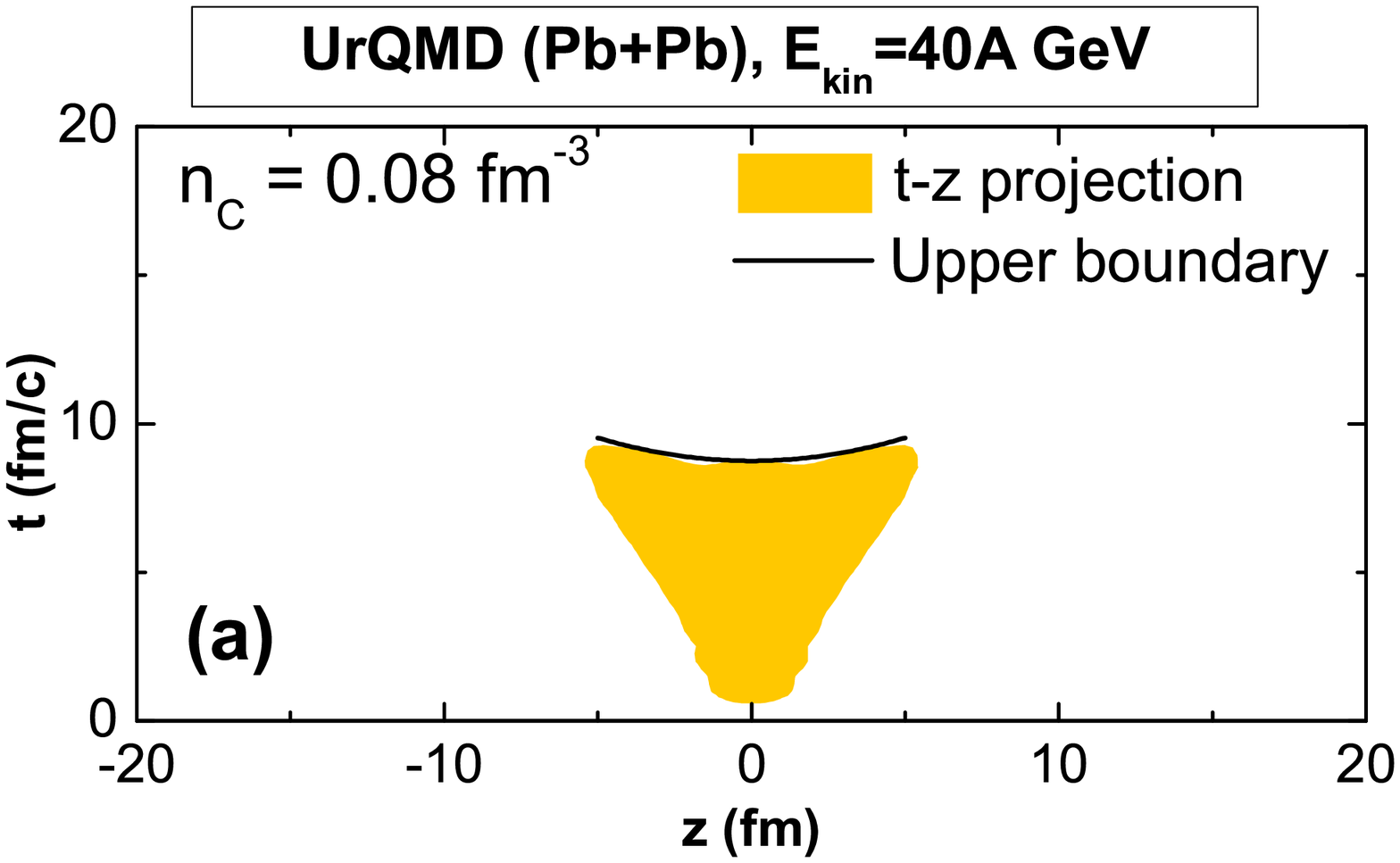}
\end{minipage}
\begin{minipage}{.48\textwidth}
\centering
\includegraphics[width=\textwidth]{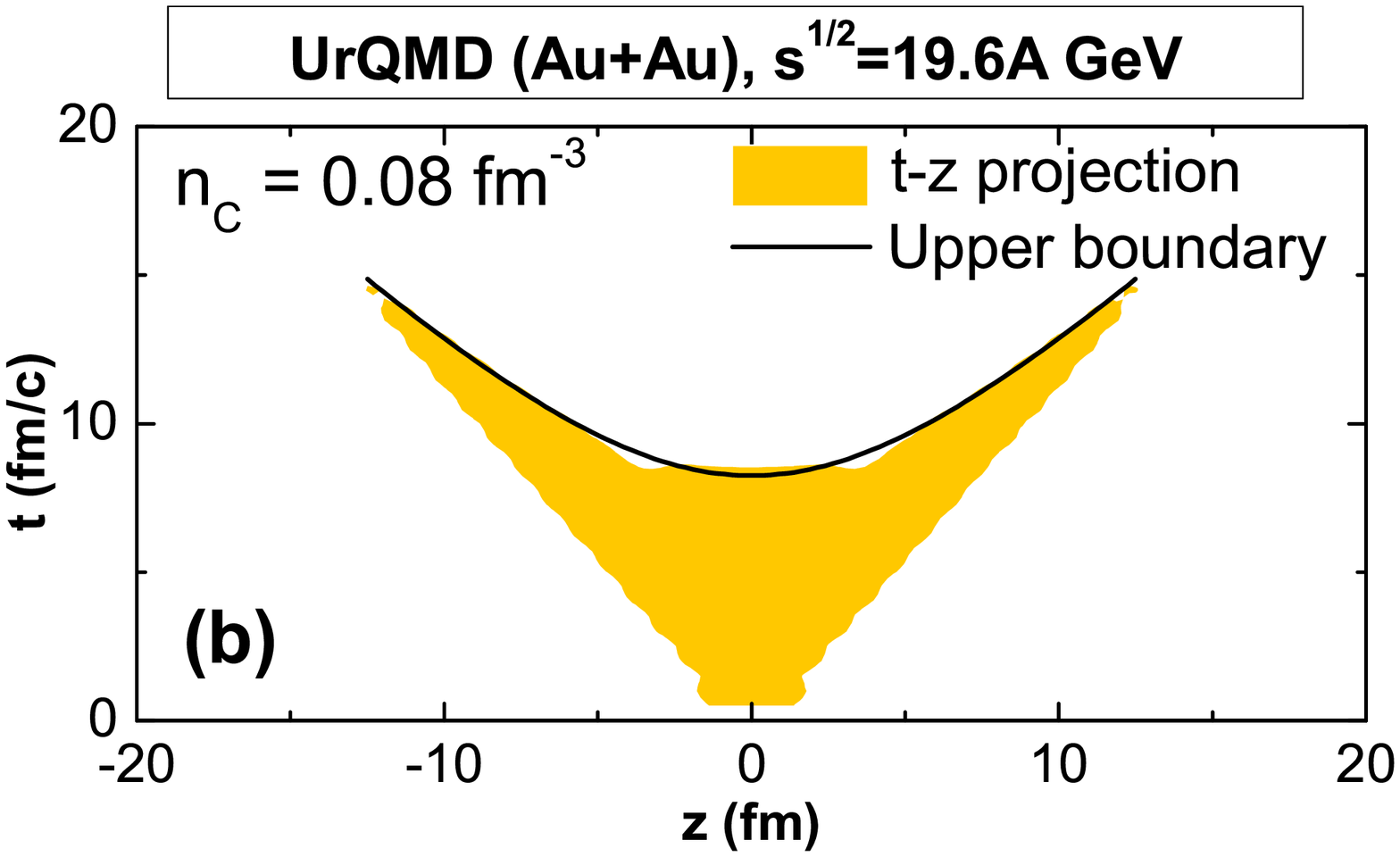}
\end{minipage}
\centering \vspace{-0.3cm}
\caption{(Color online)
The [t,z] projection of the constant $\pi^-$ density hypersurface
for (a) SPS ($E_{\rm kin}=40A$~GeV) and (b) RHIC ($\sqrt{s_{AA}} = 19.6A$~GeV) conditions.
The solid black line indicates space-like FO hypersurface boundary
approximation according to eq.~\eqref{eq:tz-appro}. }
\label{fig:Hyp-SPS-tz}
\end{figure}

Returning back to the pion FO hypersurface we reveal yet another
feature: the space-like part (time-like normal) of the $[t,\, z]$-section of
the FO hypersurface at SPS and RHIC energies can be approximated with a
$\tau_{_{\rm FO}} = {\rm const}$ hyperbola originating from possibly
different time, $t^0_{_{\rm FO}}$, than the initial time $t=0$, of the
collision (see the upper boundary curve in Fig.~\ref{fig:Hyp-SPS-tz}).
The equation for this hyperbola has the following form
\begin{equation}
t_{_{\rm FO}}(z)\ =\ t^0_{_{\rm FO}}\, +\,
\sqrt{\tau_{_{\rm FO}}^2\, +\, z^2} \,.
\label{eq:tz-appro}
\end{equation}

The fireball lifetime, $\tau$ (which is actually a fireball division time $t_{\rm d}$)
is the minimum time on the space-like FO hypersurface, i.e.,
$\tau = t_{_{\rm FO}} (z)|_{z=0}$.
Therefore, if hypersurface approximation works well for central region, then
the fireball lifetime, $\tau$, is related to hypersurface parameters as
\[ \tau \ = \  \tau_{_{\rm FO}} \, + \,  t^0_{_{\rm FO}} \,. \]

\begin{table}[!b]
 \caption{{\bf  Freeze-out hypersurface approximation parameters }} 
 \centering                                                 
 \begin{tabular}[c]{c|c|c|c|c}                              
 $E_{\rm kin}$& $\sqrt{s_{AA}}$ & $A+A$ &   $t^0_{_{\rm FO}}$ &   $\tau_{_{\rm FO}}$\\
      (A~GeV) &   (A~GeV)       &       &    (fm/$c$)      & (fm/$c$)\\
 \hline\hline                                               
 40.0 & 8.86   & $Pb+Pb$ & -7  & 15.75\\
 80.0 & 12.39  &         & -3  & 11.75\\
 158.0 & 17.32 &         & -0.75 & 9.25\\
 \hline
 202.9 & 19.6   & $Au+Au$ & -0.25 &  8.5\\
 2047.0 & 62.0   &         & -0.05 & 8.8\\
 9007.0 & 130.0  &         & 0 & 9.25\\
 \hline
 \end{tabular}
 \label{tab:tz-hyper}
\end{table}

The $[t,z]$ projections of constant $\pi^-$ density FO
hypersurface for $n_c = 0.08$ fm$^{-3}$ for SPS and RHIC energies
are depicted in Fig.~\ref{fig:Hyp-SPS-tz}.
The hypersurface boundary approximation Eq.~\ref{eq:tz-appro} is depicted by
solid black line.
The values of parameters $\tau_{_{\rm FO}}$ and $t^0_{_{\rm FO}}$ for different energies
are presented in Table~\ref{tab:tz-hyper}.
It is seen that the space-like boundary (time-like normal) is well approximated
by hyperbola of constant proper time.
Meanwhile, it turns out that
the hyperbolas originate from an
earlier times $t^0_{_{\rm FO}}<0$, than the initial time, $t=0$, of the collision.
At RHIC energies the originating time $t^0_{_{\rm FO}}$ approaches zero as the
energy rises (see Table~\ref{tab:tz-hyper}) which should be also the case
for LHC energies.

\section{Conclusions}

The present studies based in the UrQMD calculations are able to
give a reliable estimate of the final hadronic FO hypersurface
in space-time. This is a very useful information for all global estimates
and of the space-time development of the reaction. Furthermore
it gives a good guidance for the selection of
the FO hypersurface for multi-module or hybrid models, most importantly
the $[t,z]$-section of the hypersurface.
The radial section of the hypersurface is relatively simple to
model or parametrize as shown in all figures presenting central collisions.
We have to mention, however, that in peripheral collisions, where
important flow phenomena are observed (e.g. the 3rd flow component or
antiflow \cite{3rd-flow}) the FO hypersurface becomes more
involved in the transverse directions also.
Up to now the most frequently used direct FO descriptions
have simplified the FO hypersurface to $t\, =\,{\rm const.}$ or
$\tau \, = \, {\rm const.}$ hypersurfaces.
These studies indicate that the FO hypersurface is more complex, and may require
a more detailed work of finding such a surface as described in references
\cite{ChengEtal2010,PHHP12}.
We can also see that at high energies the time-like parts (space-like normals)
of the FO hypersurface are almost parallel to the light cone, therefore the
amount of particles freezing out through this part of the  FO hypersurface
is negligible compared to those crossing space-like part (time-like normal)
of the FO hypersurface.
In this work we also propose a simple alternative by observing that the
space-like part (time-like normal) of the FO hypersurface can be well
approximated with a FO-hyperbola, at  $\tau_{_{\rm FO}} = {\rm const.}$, but
originating from an earlier time  $t^0_{_{\rm FO}} < 0$, than the initial time,
$t=0$, of the collision.

In several fluid dynamical models the isochronous freeze-out or isochronous
transition from the hydro stage is assumed in hybrid models
(see, e.g., \cite{Petersen08}). These results can be improved by using an
iso-$\tau$ transition hypersurface \cite{Li09}. The optimal choice of the
proper time of the transition can be chosen based on the present
studies (see Table \ref{tab:tz-hyper}).
As discussed above up to about $\sqrt{s_{AA}} = 20$ A GeV the
origin of the freeze-out hyperbola is found to be at negative times,
which is a consequence of the string dynamics and its hadronization
in UrQMD. This work indicates that this option should also be taken into
account at the selection of freeze-out or transition hypersurface
in fluid dynamical and hybrid models.

At the same time, the minimum of the FO-hyperbola
$\tau=t_{_{\rm FO}}(z)\big|_{z=0}$
corresponds to the time moment $t=\tau$ when the many-particle
system ceases to be one unit but splits up into two separate spatial
parts which move in opposite directions from one another
with velocities approaching the speed of light with increase of collision energy.
This time $\tau$ is nothing more as a lifetime of the fireball (if we treat a
fireball as one unit) and it turns out that $\tau$ depends
so weakly on collision
energy that we can state that the fireball lifetime is approximately
invariant of the collision energy.
These results provide different possibilities to construct more realistic
models for the selection of transition or freeze out hypersurfaces.

\end{document}